# Plasmonic coupled-cavity system for enhancement of surface plasmon localization in plasmonic detectors

K J A Ooi[1,2], P Bai[1,*], M X Gu[1] and L K Ang[2,3]

[1]Plasmonics and Nanointegration Group, A*STAR Institute of High Performance Computing, 1 Fusionopolis Way, #16-16 Connexis, Singapore 138632

[2]School of Electrical and Electronic Engineering, Nanyang Technological University, Block S1, 50 Nanyang Avenue, Singapore 639798

[3]Engineering Product Development, Singapore University of Technology and Design, 20 Dover Drive, Singapore 138682

*Email: baiping@ihpc.a-star.edu.sg

**Abstract**

A plasmonic coupled-cavity system, which consists of a quarter-wave coupler cavity, a resonant Fabry-Perot detector nanocavity, and an off-resonant reflector cavity, is used to enhance the localization of surface plasmons in a plasmonic detector. The coupler cavity is designed based on transmission line theory and wavelength scaling rules in the optical regime, while the reflector cavity is derived from off-resonant resonator structures to attenuate transmission of plasmonic waves. We observed strong coupling of the cavities in simulation results, with an 86% improvement of surface plasmon localization achieved. The plasmonic coupled-cavity system may find useful applications in areas of nanoscale photodetectors, sensors, and an assortment of plasmonic-circuit devices.

## 1. Introduction

Plasmonics is the field of study of surface plasmons, the collective electron oscillations excited by coherent electromagnetic energy, which are bound to the interface between a metal and a dielectric [1,2]. The ability of plasmonic structures to squeeze light far below the diffraction limit has been harnessed to build efficient nanoscale plasmonic circuits, paving the way for on-chip integration of optics and electronics [3,4]. Great progress has been made in the development of plasmonic-circuit device components in areas of signal generation [5–7], guiding [8–10], detection [11–13] and modulation [14–16]. One of the components in plasmonic circuits is the waveguide-integrated plasmonic detector, which functions as an opto-electric transducer to interface between plasmonic waveguides and electrical devices [11–13]. When plasmonic detectors are scaled down to match the size of nanoscale electronic devices, the quantum efficiency will be dramatically reduced. Hence, the performance of a waveguide-coupled plasmonic detector will depend on resonant optical coupling and optical localization in the detector structure.

Optical coupling of a plasmonic waveguide to a plasmonic detector can be improved by using optical antennas which optimizes the energy transfer between the detector and the waveguide [17–20]. In addition, gratings or grooves are also used to improve the transmission coupling of plasmonic waves [21,22]. The plasmonic waves coupled into the detector have to be confined in the active region of the detector for effective absorption and conversion into electrical signals. The confinement can be improved by using nanocavity resonator structures [12,13,23]. In our previous works, we have applied these concepts to design an antenna-assisted Fabry-Perot nanocavity detector that strongly couples and moderately confines plasmonic waves in a germanium active material [12,13]. However, the confinement strength of Fabry-Perot resonators is limited by the refractive index difference of the cavity and its background material, governed by the Fresnel reflection equations. To solve this problem, we find insights from coupled-cavity systems present in the photonic-crystal class of devices such as the coupled-resonator optical waveguide (CROW) and coupled-cavities waveguide (CCW) [24], and recently also adapted into plasmonic systems [25,26]. The coupled-cavity technique is shown to enable efficient transmission, reflection or localization of optical power depending on how the cavities are designed. In this paper, we explore the concept of a coupled-cavity system to further improve the transmission coupling as well as localization efficiencies of our plasmonic detector.

## 2. Design theory and principles

Our proposed plasmonic coupled-cavity system is shown in Fig. 1, designed for telecommunication wavelength at 1550*nm*. It is a construct of three components: a coupler cavity, a detector cavity and a reflector cavity. Here, the plasmonic waves are guided in a hybrid plasmonic waveguide formed by a stack of aluminium-silica-silicon layered structure, with the thickness of the aluminium, silica and silicon to be 400*nm*, 150*nm* and 150*nm* respectively. These waveguide dimensions are taken from the design of plasmonic integrated-circuits which is not shown here. Directly adjacent to the hybrid plasmonic waveguide is the coupler cavity, designed as an aluminum-silica-aluminum unidirectional coupler. The detector component is a germanium Fabry-Perot nanocavity attached with a resonant monopole antenna as discussed in our previous work [13]. The reflector cavity is silica enclosed by aluminum reflectors to reflect plasmonic waves that leak out from the detector cavity. The entire ensemble is planar and 220*nm* in height, designed for CMOS standards. In the following sections, we will lay out the design principles for the coupler cavity and reflector cavity, with a brief mention of the detector cavity design.

*2.1 Coupler cavity design*

The coupler cavity is designed using the quarter-wave impedance transformer model based on transmission line theory. It improves transmission of plasmonic waves from the hybrid waveguide into the detector cavity through impedance-matching. The input impedance $Z_{in}$ (seen from the hybrid waveguide terminal), characteristic impedance $Z_0$ (coupler impedance), and output impedance $Z_L$ (for both the detector and reflector cavities) are related by the expression [27]

$$Z_{in} = Z_0 \frac{Z_L + iZ_0 \tan(\beta l)}{Z_0 + iZ_L \tan(\beta l)} \qquad (1)$$

where *i* is the imaginary constant, $\beta = 2\pi/\lambda$ the propagation constant and *l* the length of the coupler. When *l* is designed as odd quarter-multiples of wavelength, the term *tan(βl)* will tend to infinity and thus the expression can be reduced to $Z_{in}/Z_0 = Z_0/Z_L$, which is the condition for dual impedances, with a matching coupler characteristic impedance of $Z_0 = \sqrt{Z_{in}Z_L}$. Here, $Z_0$ can be optimized by selecting appropriate dielectric materials and/or by adjusting the width of the coupler-cavity.

Next, the real resonance lengths of the cuboid coupler rod (cross-section of 150nm × 220nm) shown in Fig.1 can be obtained by applying optical regime effective wavelength scaling rules. However, as there is no simple analytical equation to describe a cuboid coupler rod, we will approximate it with Novotny's equation of the resonant optical response for a cylindrical coupler rod in a homogenous background medium [28]:

$$\lambda_{eff} = \frac{\lambda}{\sqrt{\varepsilon_s}} \sqrt{\frac{x}{1+x}} - \frac{4r}{N} \quad (2a)$$

$$\text{and } x = 4\pi^2 \varepsilon_s \left(\frac{r^2}{\lambda^2}\right) \left[ 13.74 - 0.12 \left(\frac{\varepsilon_\infty + 141.04 \varepsilon_s}{\varepsilon_s}\right) + 0.12 \left(\frac{\sqrt{\varepsilon_\infty + 141.04 \varepsilon_s}}{\varepsilon_s}\right) \frac{\lambda}{\lambda_p} \right]^2 \quad (2b)$$

where $r$ is the cylindrical coupler rod radius, $\lambda_p$ the plasma wavelength, $N$ the order of resonance, $\varepsilon_s$ the permittivity of dielectric and $\varepsilon_\infty = \varepsilon_m$ ($\lambda \rightarrow 0$) the infinite frequency limit of dielectric function of metal $\varepsilon_m$. Here, to obtain good approximation of $r$ for the cuboid coupler rod, we divide the rectangular cross-section into eight radial segments in a "Union Jack" fashion, and then average the lengths of the diagonals and perpendiculars. We obtain $r \approx 110nm$ using this method. By using material values of aluminium ($\varepsilon_m = -252.5+46.07i$, $\lambda_p = 96.7nm$ and $\varepsilon_\infty = 1$) and silica ($\varepsilon = 2.085$) in Ref. [29], we get the first resonant cuboid coupler length at 159nm.

*2.2 Reflector cavity design*

The reflector cavity suppresses out-coupling of plasmonic waves. It has to be designed as an off-resonant destructive-interfering cavity in order to attenuate any build-up of optical power inside the cavity. The attenuation depends on the relative phase-shift on reflection, $\Delta\phi$, and the phase-shift on propagation, $\beta$, between the two reflective interfaces. Destructive interference occurs when the cavity length $d$ fulfills the condition:

$$2d\beta + \Delta\phi = (2n-1)\pi, \quad n = 1,2,3... \quad (3)$$

Thus to determine the cavity length, the $\Delta\phi$ between the germanium-silica and the silica-aluminum interfaces has to be calculated. The Fresnel reflection coefficient of an interface is given by:

$$r = \frac{n_s - (n-ik)}{n_s + (n-ik)} = \text{Re}(r) + \text{Im}(r) \quad (4)$$

where $n_s$ is the refractive index of the incident medium and $n-ik$ the refractive index of the reflector. The phase-shift on reflection $\phi$ is given by:

$$\phi(r) = \tan^{-1} \frac{\text{Im}(r)}{\text{Re}(r)} = \frac{2n_s k}{n_s^2 - n^2 - k^2} \quad (5)$$

Using $n_s = 1.44$ for silica, $n+ik = 4.276+0.0242i$ for germanium and $n+ik = 1.44+16i$ aluminum, $\phi$ for germanium and aluminum are worked out to be $0.943\pi$ and $0.983\pi$ respectively, with a $\Delta\phi$ of $0.04\pi$. The first off-resonant cavity length can thus be estimated around $d = 258nm$.

*2.3 Detector cavity design*

The principles of the detector cavity have already been discussed in our previous work [13]. A monopole antenna, with a resonant length of $600nm$, is directly mounted onto the hybrid plasmonic waveguide terminal. Sandwiched in between the antenna and the waveguide is a germanium Fabry-Perot resonator, with a resonant length of $150nm$ and width of $60nm$. The monopole antenna strongly couples and concentrates the plasmonic waves into the germanium nanocavity, while the Fabry-Perot resonator reasonably confines the plasmonic waves for optical power absorption in the germanium active material.

## 3. Results and Discussion

We present some simulation results of the plasmonic coupled-cavities simulated with the transient solver in C. S. T. Microwave Studio 2010. In our simulations, the material parameters are set as: aluminium ($\varepsilon_m = -252.5+46.07i$, $\lambda_p = 96.7nm$ and $\varepsilon_\infty = 1$), silicon ($\varepsilon = 12.11$), silica ($\varepsilon = 2.085$) and germanium ($\varepsilon = 18.28 + 0.0485i$) which are found in Ref. [29]. The simulation domain is embedded into a silica background substrate. A $1550nm$ wavelength signal is then injected into the frontend of the hybrid waveguide using a waveguide-port.

Fig. 2(a) shows the optical localization efficiency for different coupler lengths, resonant at $150nm$ and $650nm$. It is noted that the simulated resonant lengths slightly differ from the analytical lengths due to an inhomogeneous background interfacing material, and the use of a cuboid coupler rod instead of a cylindrical one. The second resonance is weaker than the first resonance due to additional propagating losses incurred by a longer coupler. To clearly show the role of the coupler in enhancing transmission coupling and separate effects from the antenna, we simulate isolated coupler-cavities of various lengths and show their current density maps in Fig. 3. It is observed that the resonant couplers (150nm and 650nm) have their current densities concentrated at the coupler ends (Fig. 3(a), (c)). On the other hand, the off-resonant coupler (400nm) has its current density spread along the coupler arm (Fig. 3(b)). These observations are consistent with theoretical predictions.

To investigate the role of the reflector, we examine in Fig. 2(b) the in-coupling efficiency and out-coupling efficiency of both cases with and without an optimized reflector ($250nm$ in length and $650nm$ in width). The schematic of both cases are depicted in Fig. 2(c). The in-coupling efficiency is defined as the optical transmission at the coupler-detector cavity interface, while the out-coupling efficiency defined as the optical transmission at the detector-reflector cavity interface. We find that the out-coupling efficiency is significantly reduced by using an optimized reflector, thus verifying the role of the reflector to attenuate leakage of plasmonic waves from the detector cavity.

Meanwhile, in the examination of the in-coupling efficiency in Fig. 2(b), we found an interesting observation. We see that in the presence of an optimized reflector, the crests for the in-coupling efficiency are inflated while the trough is depressed, indicating that the reflector plays an important role in strengthening the coupler resonance as well. This shows that the three cavities formed a very strongly-coupled system: the optical localization strength of the system *as a whole is greater than the sum of* its individual cavities' resonance strengths.

The performance of the coupler also depends on its characteristic impedance, which in turn depends on the refractive index and the width of the dielectric material in the coupler-cavity. Here, we limit our discussion to the width of the coupler cavity, as the dielectric material used for the coupler cavity is silica for ease of fabrication. As shown in Fig. 4, optical coupling reaches a maximum when coupler width is around 170*nm*. Further, optical coupling decreases by merely 2% when the width of the coupler is changed by ±30*nm* from the optimum value i.e. 170*nm*. Here, it should be noted that setting the coupler width at 170*nm* will introduce "kinks" into the device structure, noticeable from the schematic diagram in the inset of Fig. 4. From fabrication view point, a streamlined geometry is preferred. Hence, coupler width can be set at 150*nm* with negligible degradation of the device performance.

Fig. 5(a) shows the optical localization efficiency as a function of reflector cavity dimensions. The maximum attenuation occurs at cavity lengths 250*nm* and 800*nm*, which is consistent with analytical predictions. The cavity width, meanwhile, is not critical as long as it falls in the range of 200*nm*–700*nm*. Simulation results in Fig. 5(b) revealed that outside this range, there is significant radiation loss from the reflector cavity occurring through the encapsulating substrate in the z-direction. This implies that the cavity width controls the transverse mode-matching of the plasmonic waves with the reflector [30].

For practical circuit applications, the reflectors can also be used as electrodes to apply a bias voltage across the detector cavity. For this purpose, a narrow isolation cut through the reflector is needed for electrical isolation purposes (labeled in Fig. 1). This isolation layer does not reduce the reflective properties of the reflector as long as its width is below 300*nm*.

Finally, we compare the performance of the coupled-cavity system with a lone nanocavity resonator system, the latter at a benchmarked optical absorption efficiency figure of 42% [13]. The coupled-cavity system is found to have an optical absorption efficiency figure of 78%, representing an increase of 86%.

## 4. Conclusion

In summary, we have designed a plasmonic coupled-cavity system that strongly localizes surface plasmons in a detector nanocavity by employing a strongly coupled coupler-detector-reflector cavity system. Optical in-coupling is increased with the use of a resonant quarter-wave coupler cavity, while out-coupling is attenuated by an off-resonant reflector cavity. The coupled-cavity system improves the optical absorption efficiency of a lone nanocavity resonator system by 86%. The proposed coupled-cavity system may also find useful applications in plasmonic devices other than the detector, which includes switches, modulators, emitters and sensors.

**Acknowledgments**

This research is supported by A*STAR SERC Grant, Metamaterials Programme: Nanoplasmonics Grant No. 092 154 0098. One of us (KJAO) is supported by a PhD scholarship funded by the MOE Tier2 grant (2008-T2-01-033).

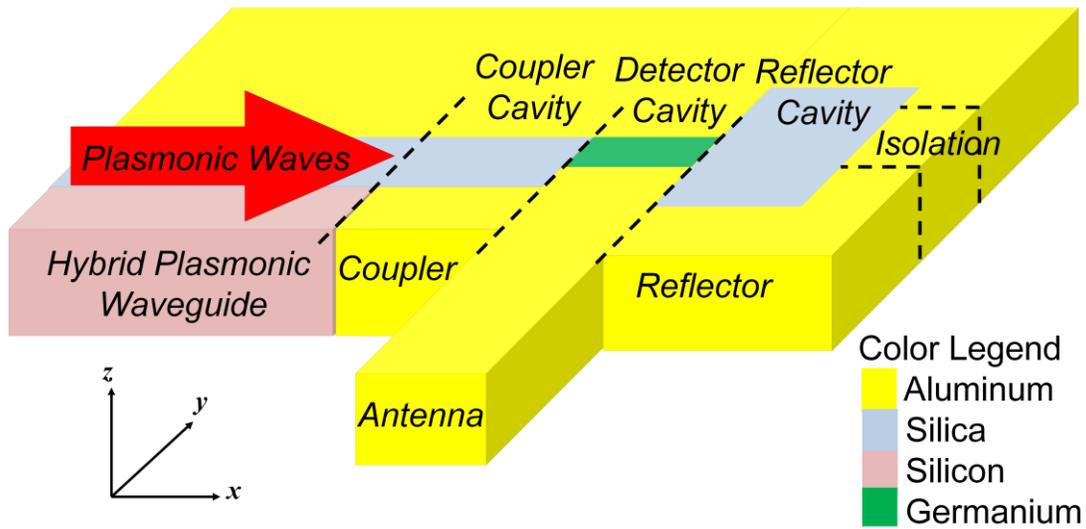

Fig. 1. Schematic of a coupler-detector-reflector coupled-cavity system.

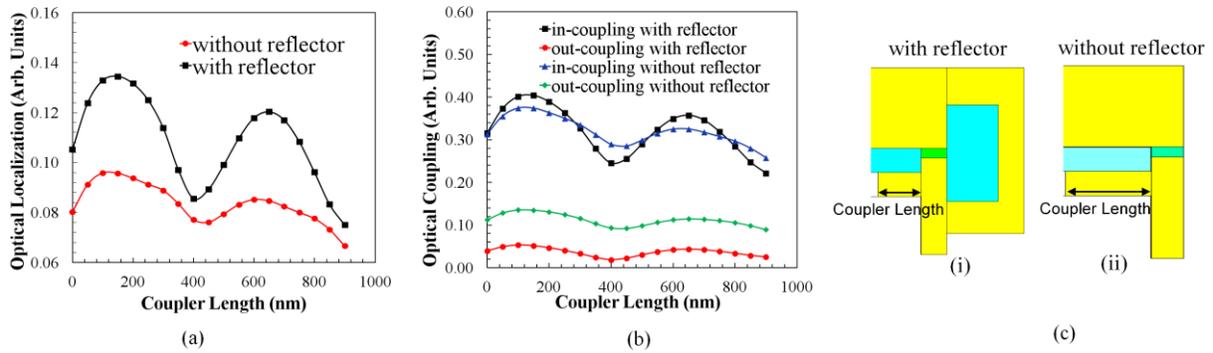

Fig. 2. (a) Optical localization efficiency dependence on coupler lengths with and without reflectors. (b) In-coupling and out-coupling efficiency dependence on coupler lengths with and without reflectors. The presence of the reflector attenuates the optical out-coupling and increases the optical in-coupling contrast between the on-resonant and off-resonant coupler. (c) Schematic of coupled-cavity system (i) with and (ii) without reflectors.

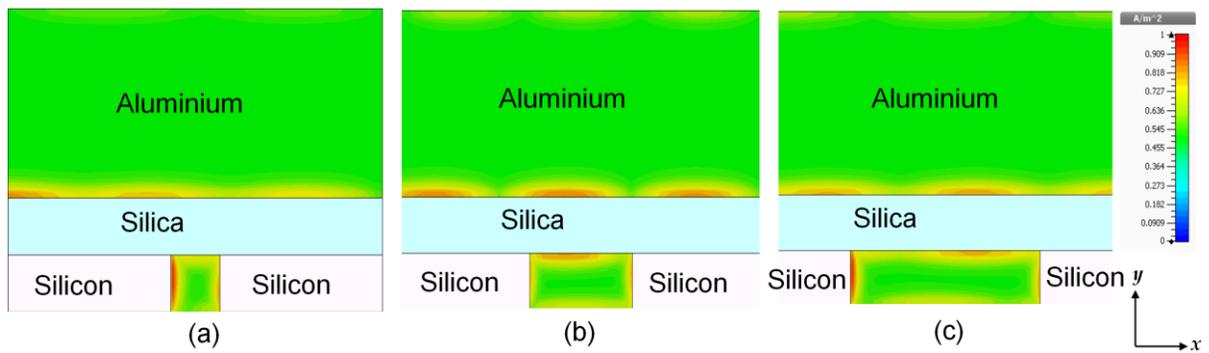

Fig. 3. Current density maps across the center of system in the xy plane for isolated couplers of length (a) 150*nm*, (b) 400*nm*, and (c) 650*nm*, which correspond to first maxima, second minima, and third maxima in Fig 2(a), respectively.

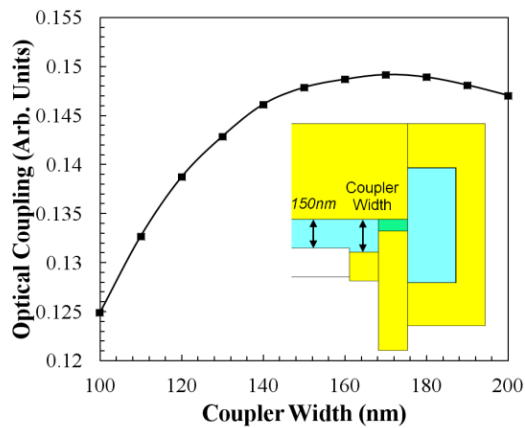

Fig. 4. Optical coupling dependence on coupler-cavity width. Inset: schematic showing the coupler-cavity width.

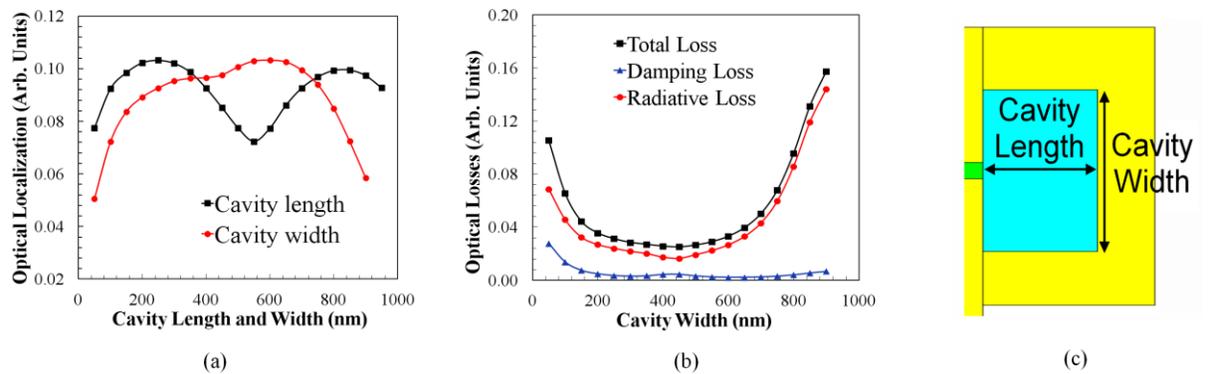

Fig. 5. (a) Optical localization efficiency dependence on reflector cavity lengths and widths. (b) Total, damping and radiative optical losses from the reflector cavity. (c) Schematic of the reflector cavity.